\documentclass[floatfix,showpacs,preprintnumbers,amsmath,amssymb,article]{revtex4}

\usepackage{graphics}
\usepackage{graphicx}
\usepackage{dcolumn}
\usepackage{bm}
\usepackage{multirow}
\usepackage{amssymb}
\usepackage{perpage}	 

\begin{document}

\title{Magnetic Field Probing of an SU(4) Kondo Resonance in a Single Atom Transistor}
\author{G.C. Tettamanzi$^{1,2}$}
\email{g.tettamanzi@unsw.edu.au}
\author{J. Verduijn$^{1,2}$}
\author{G.P. Lansbergen$^{1}$}
\author{M. Blaauboer$^{1}$}
\author{M.J. Calder\'on$^{3}$}
\author{R. Aguado$^{3}$}
\author{S. Rogge$^{1,2}$}
\affiliation{$^{1}$Delft University of Technology, Kavli Institute of Nanoscience, Department of Quantum Nanoscience, Lorentzweg 1, 2628 CJ Delft, The Netherlands}   
\affiliation{$^{2}$School of Physics, University of New South Wales, Sydney NSW 2052, Australia}
\affiliation{$^{3}$Instituto de Ciencia de Materiales de Madrid, ICMM-CSIC Cantoblanco, E-28049 Madrid, Spain}

\date{\today}

\begin{abstract}
Semiconductor devices have been scaled to the point that transport can be dominated by only a single dopant atom. As a result, in a Si Fin Field Effect Transistor Kondo physics can govern transport when one electron is bound to the single dopant. Orbital (valley) degrees of freedom, apart from the standard spin, strongly modify the Kondo effect in such systems. Owing to the small size and the s-like orbital symmetry of the ground state of the dopant, these orbital degrees of freedom do not couple to external magnetic fields which allows to tune the symmetry of the Kondo effect. Here we study this tunable Kondo effect and demonstrate experimentally a symmetry crossover from a SU(4) ground state to a pure orbital SU(2) ground state as a function of magnetic field. Our claim is supported by theoretical calculations that unambiguously show that the SU(2) symmetric case corresponds to a pure valley Kondo effect of fully polarized electrons.

\end{abstract}

 \maketitle

The resistance of metals with magnetic impurities anomalously increases as one decreases the temperature. This Kondo effect \cite{Hewson} can be explained as the screening of the localized spin of the magnetic impurity by the spins of the de-localized electrons in the metal. As a consequence of this screening, the localized spin and the itinerant ones form a many-body singlet with binding energy $T_K$, which defines the low temperature scale at which Kondo physics appears. A few years ago, it was shown that quantum dots (QDs)~\cite{Cro540} behave as Kondo impurities. The transport properties of QDs in the Kondo regime are quite remarkable: starting from an insulating QD in the Coulomb blockade regime at high temperatures,  the linear conductance reaches the maximum unitary value of a perfect quantum conductor, namely $G\equiv dI/dV_b|_{V_b\rightarrow 0}=2e^2/h$ as the temperature is reduced well below $T_K$~\cite{Hewson}. At finite bias voltages $V_b$, Kondo physics manifests as a zero-bias anomaly in the $dI/dV_b$ curves whose width is roughly given by $T_K$. The Kondo effect in QDs originates from quantum fluctuations of the charge residing in the QD: electrons can transit through virtual states on a time-scale which is shorter than allowed by the Heisenberg uncertainty principle~\cite{Hewson}. This mechanism generates effective spin flips which in turn lead to Kondo physics. Importantly, the role of the electron spin can be replaced by any other quantum degree of freedom such as e.g. orbital momentum \cite{Bor026602,Zar2043,Jar484,Shi195345,Lim205119,Cho067204,Lan455}, giving rise to exotic Kondo effects. Furthermore, the simultaneous presence of both a spin- and an orbital- degeneracy leads to an SU(4)- Kondo effect, where SU(4) refers to the symmetry of the corresponding Kondo ground state~\cite{Bor026602,Zar2043,Jar484, Shi195345, Lim205119,Cho067204,Lan455}.

In the past, SU(4) Kondo symmetry has been predicted to arise in parallel double quantum dot systems \cite{Zar2043}, but so far it has only been clearly observed in carbon nanotubes~\cite{Jar484} and in single dopant devices in Si ~\cite{Lan455}. Si is a good candidate for observing SU(4) Kondo physics due to its six-fold valley (orbital) degeneracy of the conduction band and orbital effects are the most probable cause of the unexplained (at the time) behaviours observed in early studies of Kondo in Si QDs \cite{RokR16319}. Here we show a fully tunable Kondo effect in a Si Fin Field Effect Transistors (FinFETs)~\cite{Sel206805,Lan656, Pie133,Gol075401} [see inset of Fig. 1(a)]. We use a gate voltage to bring the orbital states into degeneracy, while independently tuning the spin splitting by means of an external magnetic field. Interestingly, in our system, orbital Kondo physics survives at very high magnetic fields (even for $\Delta_Z \gtrsim k_{B}T_{K}$, with  $\Delta_Z\equiv g^{*}\mu _{B}$B being the spin splitting) allowing us to tune our device from an SU(4) to an SU(2) symmetry. In this latter case, our results can be understood as a pure orbital Kondo effect of spin-polarized electrons \cite{Bor026602,Tro125323}. Our claim is fully supported by theoretical calculations that take into account both thermal and quantum fluctuations in a non-pertubative way.%

Single dopants can be individually addressable in FinFETs~\cite{Sel206805,Lan656,Pie133,Gol075401}. Our devices consist of Si nanowire connected to large contacts etched in a $60$ nm layer of p-type Silicon On Insulator (SOI). The wire is covered with nitride oxide ($1.4$ nm equivalent SiO$_{2}$ thickness) and a narrow poly-crystalline silicon wire is deposited perpendicularly on top to form a gate on three faces. Doping by ion implantation with As over the entire surface forms n-type degenerate source, drain and gate electrodes while the channel protected by the gate remains p-type. The conventional operation of this n-p-n field effect transistor is to apply a positive gate voltage to create an inversion in the channel and allow a current to flow. Unintentional As donors may be present below the Si/insulator interface showing up in the sub-threshold transport characteristics~\cite{Sel206805,Lan656}. 

\begin{figure*}
\begin{center}
\resizebox{9 cm}{!}{\includegraphics{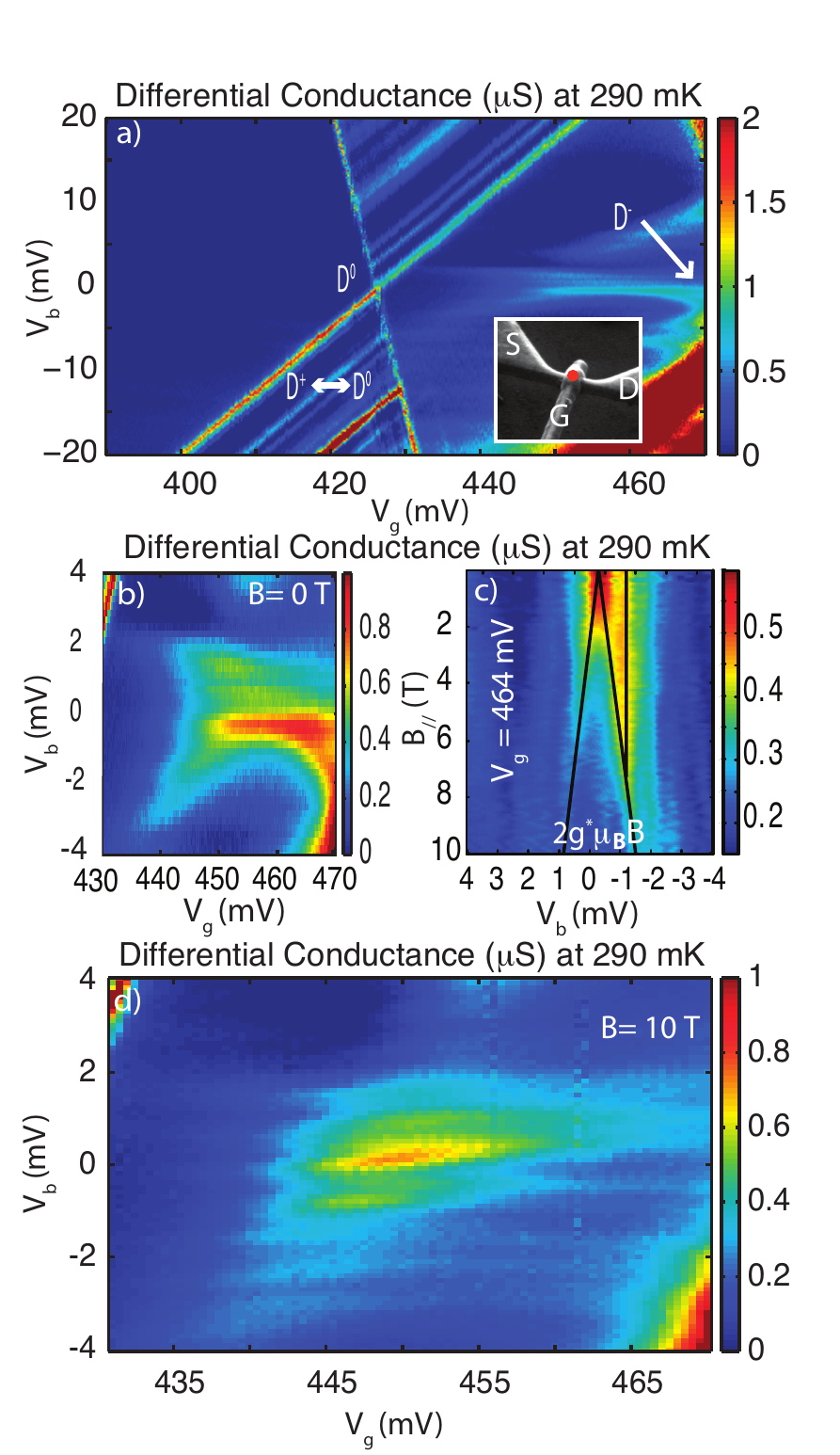}} 
\caption{a) Stability diagram of the differential conductance (for \textit{B} = 0 T and $T$= 290 mK) of a single atom transistor. The transport resonances inside the diamond, which are Kondo related, should not be allowed due to Coulomb blockade (CB). In the inset, a Scanning Electron Micrograph (SEM) of our device with the schematic position of the dopant in the channel (in red), is shown. b) \textit{Enlargement of the Kondo region of Fig. \ref{fg:DH64ChipC2T300mKStabCONDBIS}} (a). c) Evolution under magnetic field of the Kondo resonance at 290 mK and for $V_g$= 464 mV (far from the degeneracy point). For this gate voltage value, the expected 2g$\mu_{B}$B splitting of the central Kondo resonance is observed and as expected for non-degenerate SU(4) Kondo, the side resonance do not split in magnetic field \cite{Lan455}. d) Same as Fig. \ref{fg:DH64ChipC2T300mKStabCONDBIS} (b) but for \textit{B}= 10 T. In this figure it is possible to observe an unusual (compared to other Kondo regions) survival of the zero bias peak in the $V_g$ $\approx$ 450 mV region.}
\label{fg:DH64ChipC2T300mKStabCONDBIS}
\end{center}
\end{figure*}

From a large ensemble of FinFET devices, we select those that show the fingerprint of single donor transport, which essentially consists of a pair of characteristic transport resonances associated with the one-electron ($D^{0}$) and two-electron ($D^{-}$) charge states of the single donor~\cite{Sel206805,Lan656,Pie133}. The electron ground state originates from the hybridisation of the donor hydrogen-like state with a quantum well state formed at the Si/insulator interface by the high electric field in the channel~\cite{martins04,calderonPRL06,Sel206805, Lan656}. Valley degeneracy in Si is strongly broken at the donor (valley splitting $\sim 20$ meV \cite{Wil1068}) but a nearly two-fold degeneracy remains at the interface. As a result, the two lower orbitals show a splitting $\Delta$ typically on the order of a few meV~\cite{Lan656}. This valley splitting may be modified externally by applying a gate voltage~\cite{Lan656} which affects interface quantum well depth and the transparency of the barriers leading to a modification of the energy levels of the hybridized electron ground state wave-function  \cite{Gei2061,Sim804}. On the other hand, the s-like orbital degree of freedom is still dominant in this system \cite{Lan656}, and, consequently, the valley splitting is virtually independent to magnetic fields applied parallel to the channel.

\begin{figure*}
\begin{center}
\resizebox{8 cm}{!}{\includegraphics{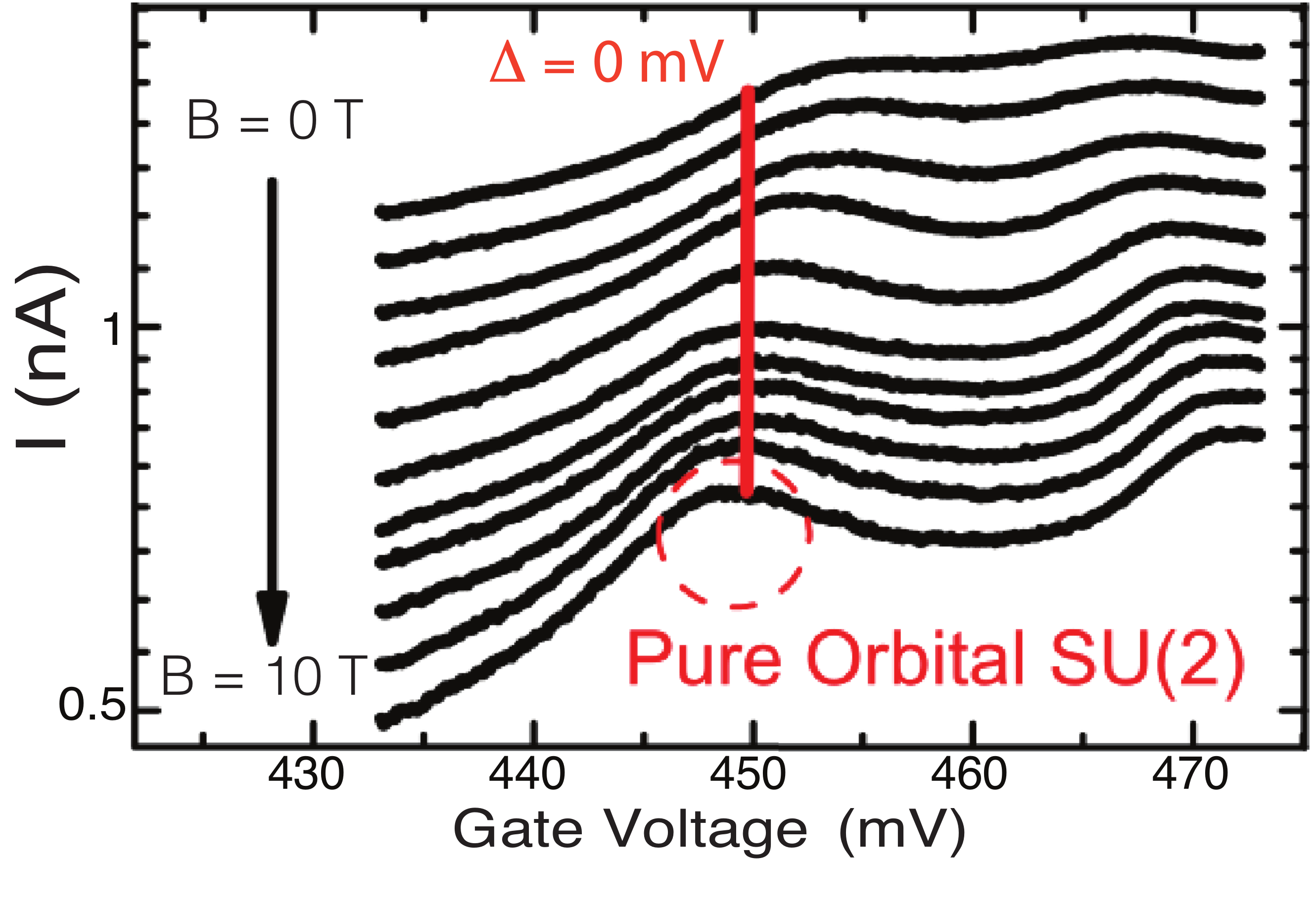}}
\caption{Current measured in the CB region of Fig. \ref{fg:DH64ChipC2T300mKStabCONDBIS} (a) for zero bias voltage, $T$ = 288 mK and increasing values of $B$. The peak observed for $B$ = 10 T and around $V_{g}$ $\sim$ 450 mV (red circle) is related to pure orbital fluctuations while the red line indicates the crossover from a Kondo resonance with SU(4) symmetry and degenerate levels to a pure SU(2) orbital Kondo. The Kondo temperatures for $B$ = 0 T and $V_g$ = 450 $mV$, 455 $mV$ and 464 $mV$ have been estimated to be $\approx$ 8 K, 6 K and 4 K respectively. An increasing offset has been added to the curves for clarity.}
\label{fg:H64C2T288mKkondoRegionvsBfinalA}
\end{center}
\end{figure*}

Transport spectroscopy measurements as a function of gate and bias voltage have been performed using a lock-in technique, typically with a 50 $\mu$V amplitude ac voltage, and a $^3He$ cryostat with a base temperature of $270$ mK. A magnetic field $B$ parallel to the direction of the channel of the FinFETs has also been applied. All the data presented here correspond to a device in which a single dopant atom can be individually addressed due to strong capacitive coupling with the source and drain contacts~\cite{Sel206805} (see Fig. \ref{fg:KONDOTat450mVand10T} in the supplementary section).

\begin{figure*}
\begin{center}
\resizebox{8 cm}{!}{\includegraphics{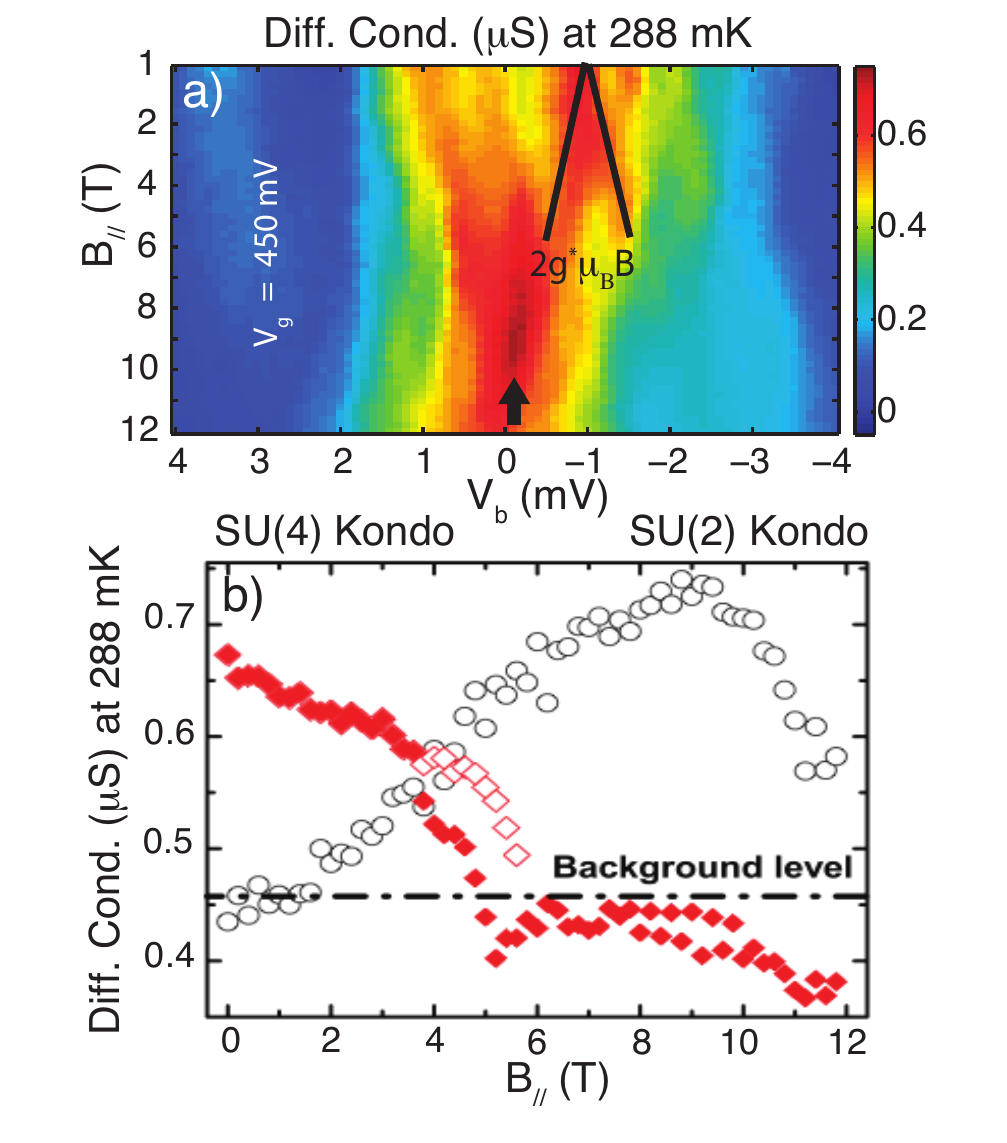}}
\caption{a) Evolution of the Kondo peak at 288 mK and for $V_{g}$ = 450 mV. For this gate voltage value no splitting is observed for $B$ $\gtrsim$ 4 T as the central peak survive (black arrow) also for very high magnetic fields. b) Traces that show the evolution under increasing magnetic field of the peaks height at $V_{b}$ $\approx$ -1 mV (rhombuses) and of the peak at $V_{b}$ = 0 mV (circles).}
\label{fg:H64C2KONDOCOND288mK0T10TA}
\end{center}
\end{figure*}

In Fig. \ref{fg:DH64ChipC2T300mKStabCONDBIS} (a) the differential conductance \textit{G}$=dI/dV_{b}$ of our device as a function of the gate $V_g$ and bias $V_b$ voltage is shown at $T$= 290 mK and $B$ = 0 T. A Coulomb diamond is defined  between the $D^0$ peak at $V_g \simeq 426$ mV and the $D^-$ peak at $V_g \simeq 480$ mV~\cite{Sel206805, Lan656}. As it has been already shown in a previous publication~\cite{Lan455}, for our system the Kondo resonances around $V_b \sim 0$ have the typical SU(4) Kondo signature. In particular, we observe a split Kondo resonance associated to fluctuations in two quasi-degenerate orbital states and a central one associated with spin fluctuations. This is illustrated in the blow up in Fig. \ref{fg:DH64ChipC2T300mKStabCONDBIS} (b) which reveals a non-zero bias central peak and side peaks, all of these scaling in temperature according to the same empirical law \cite{Gol5225}, as described in Ref.~\cite{Lan455}. The spin-related peak occurs at non-zero bias due to asymmetries in the contacts \cite{Sim804,Lan3} (see also theory section in the supplementary section) and to the possible presence of Fano physics \cite{Liu135226,Ald1}, the signature of which has been already observed in our systems \cite{Ver072110}, while the side peak arises due to the valley splitting ($\Delta \neq$ 0 mV) of the donor bound electron ground state. The evolution of the Kondo peaks with an increasing magnetic field for $V_g \simeq 464$ mV is shown in Fig. \ref{fg:DH64ChipC2T300mKStabCONDBIS} (c). The strongest peak splits $2 g^* \mu_B B$, as expected \cite{Cro540}, while the satellite valley peak remains unaffected by the magnetic field \cite{Lan455}. These features are common for most values of the gate voltage~\cite{Lan455}. The behaviour with magnetic field is dramatically different for $V_g \sim 450$ mV (see Fig. \ref{fg:DH64ChipC2T300mKStabCONDBIS} (d), Fig. \ref{fg:H64C2T288mKkondoRegionvsBfinalA} and Fig. \ref{fg:H64C2KONDOCOND288mK0T10TA} (a)). This gate voltage is special as it corresponds to a valley degeneracy point ($\Delta$ = 0 mV)~\cite{Lan455}. At gate voltages $V_g$ = 450 mV, 455 mV and 464 mV, the splitting of the orbitals progressively increases, which results in a decrease of the corresponding Kondo temperature \cite{Zar2043,Eto95}, i.e:. $T_K$ = 8 K, 6 K and 4 K respectively. Therefore, for $V_g \sim 450$ mV, we have a four-fold degeneracy that reflects in a SU(4) Kondo effect. In the absence of a magnetic field, the dominating resonance is a structured peak at non-zero bias: $V_b$ = -1 mV. However, as a magnetic field is applied, this non-zero bias resonance suffers a 2g$\mu_{B}$B splitting and gets suppressed while a simple central ($V_b\sim 0$ mV) peak gets enhanced and dominates at large fields $B \gtrsim 4 $~T. These behaviours are depicted in Fig. \ref{fg:H64C2KONDOCOND288mK0T10TA} (a), with the differential conductance as a function of $V_b$ and the magnetic field, and in Fig. \ref{fg:H64C2KONDOCOND288mK0T10TA} (b) with the evolution of the heights of the two peaks, at $V_b\simeq -1$~mV and $V_b\simeq 0$~mV, as a function of the magnetic field. From these data we can observe that, for $B=10$~$T$, a central resonance (with $T_{K}$ $\approx$ 6 K, see Fig. 5 of supplementary section) dominates the stability diagram (as also shown in Fig. \ref{fg:DH64ChipC2T300mKStabCONDBIS} (d)). The development of this central ($V_b \simeq 0$) peak with increasing magnetic field is also illustrated in Fig. \ref{fg:H64C2T288mKkondoRegionvsBfinalA}, where the current versus $V_g$ is plotted for different values of $B$. As the black arrow in Fig. \ref{fg:H64C2KONDOCOND288mK0T10TA} (a) indicates, this peak does not shift or split in $V_b$ when increasingly high values of $B$ are applied, implying that it is related to the orbital (valley) degree of freedom which must be preserved during tunnelling \cite{Lan455,Shi195345}.

\begin{figure*}
\begin{center}
\resizebox{10 cm}{!}{\includegraphics{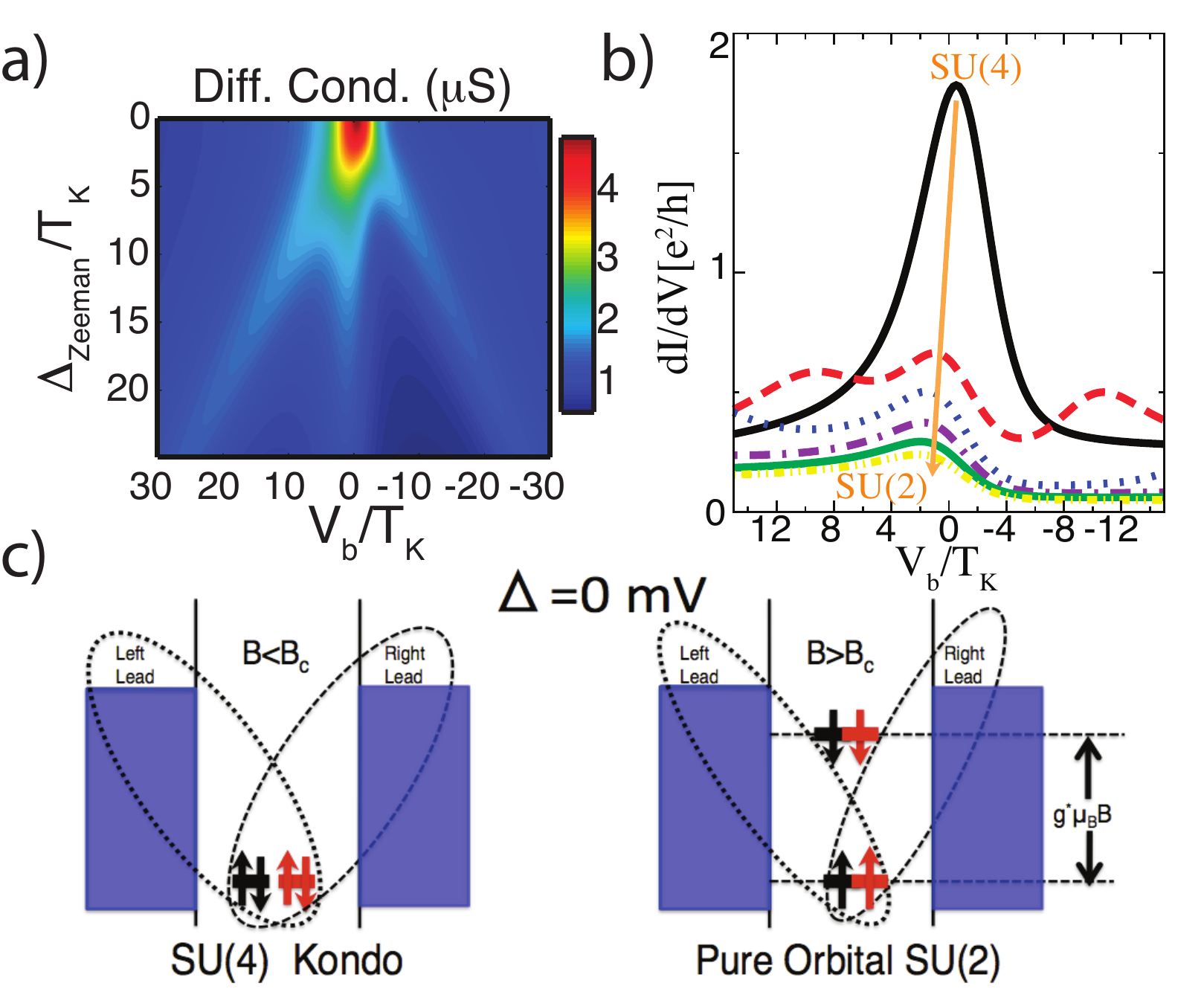}}
\caption{a) Theoretical investigation of the degenerate SU(4) to pure orbital SU(2) crossover as schematically described in Fig. \ref{fg:H64C2KONDOCOND288mK0T10TB} (c), performed using the non-crossing approximation (see supplementary section and Ref.~\cite{Lim205119,Cho067204}). b) $G$ versus bias traces for increasing values of $B$ along the crossover as in Fig. \ref{fg:H64C2KONDOCOND288mK0T10TB} (a). The black, red, blue, violet, green and yellow curves represent respectively the $\Delta_{Zeeman}$ = 0 $T_{K}$, 10 $T_{K}$, 20 $T_{K}$, 30 $T_{K}$, 40 $T_{K}$ and 50 $T_{K}$ cases. c) Schematic of the phenomenom discussed in this paper. Starting from degenerate valleys (represented in black and in red in this figure), an increasing magnetic field $B$ saturates the spin degree of freedom due to the Zeeman splitting, producing a crossover from an SU(4) Kondo symmetry to a SU(2) pure orbital Kondo symmetry.}
\label{fg:H64C2KONDOCOND288mK0T10TB}
\end{center}
\end{figure*}

 This observation is of importance as the conservation of valley index upon tunnelling is most probably sample dependent and still the subject of theoretical debate \cite{Shi195345,Lan455}. Moreover, the large value of the magnetic field indicates the spin must be completely polarized \cite{Bor026602,Tro125323}, such that the only Kondo active degree of freedom is the orbital one (implying an SU(2) symmetry as schematically shown in Fig. \ref{fg:H64C2KONDOCOND288mK0T10TB} (c)). Last, it is possible to interpret the white circles in Fig. \ref{fg:H64C2KONDOCOND288mK0T10TA} (b) as $G$ versus $T/T_K$ (as $B$ increases, $T_K$ reduces so $T/T_K$ increases). The observed $G$ versus $T/T_K$ dependence indeed follows the non-monotonic behaviour expected for a two-stage Kondo system, in agreement with the theoretical description of the SU(4) to SU(2) symmetry crossover \cite{Shi195345,Lim205119,Eto95}.

Having discussed the main experimental features, we now focus on a theoretical analysis that supports our previous interpretations. Our theoretical calculations are performed by considering a QD with two (nearly) degenerate localized orbitals (valleys) coupled to reservoirs. The effect of the external magnetic field is to lift the spin degeneracy of the single-particle energy levels, with the resulting splitting denoted $\Delta_{Zeeman}$ (see Fig. \ref{fg:H64C2KONDOCOND288mK0T10TB} (a) and \ref{fg:H64C2KONDOCOND288mK0T10TB} (b)).  The final results are obtained by using the so-called non-crossing approximation (NCA), a fully non-perturbative theory which includes both thermal and quantum fluctuations, complemented with the Keldysh-Green's function method to take into account non-equilibrium effects, due to the finite bias voltage $V_b$. More details on the theory can be found in the supplementary section. In Fig. \ref{fg:H64C2KONDOCOND288mK0T10TB} (a) and \ref{fg:H64C2KONDOCOND288mK0T10TB} (b), the calculated differential conductance as a function of $V_b$ and the magnetic field is plotted. At zero magnetic field, we obtain a SU(4) Kondo resonance which is maximum at non-zero bias voltage. This effect is attributed to an asymmetry in the way the dopant is coupled to the reservoir \cite{Sim804}. We treat the case of orbital degeneracy, namely $\Delta$ = 0 mV. As a magnetic field is applied, this non-zero bias resonance splits in three: the two outer peaks correspond to inelastic cotunnelling-like processes in which spin flips occur mediated by Kondo fluctuations in the orbital degree of freedom. The central peak corresponds to orbital Kondo processes in which the spin is conserved. As the magnetic field increases, the central peak moves towards zero bias and the outer peaks have no longer the same height (the one corresponding to positive voltages being higher). This, again, can be easily understood as resulting from coupling asymmetry: for positive voltages, the distance between the spin-polarized orbital dopant levels and the left chemical potential decreases resulting in a stronger Kondo effect as compared with the one for negative voltages where the right reservoir (the one with lower chemical potential) is less coupled to the dopant levels. Importantly, only quantum fluctuations between orbital levels are active so the underlying symmetry of the Kondo effect is SU(2). We have assumed that the valley index is preserved during tunnelling. This is a necessary condition for the observation of the SU(2) orbital Kondo effect as valley index mixing would destroy it \cite{Cho067204}. It is important to mention that, in order to obtain pure orbital SU(2) physics in the theoretical calculations, we need to include higher magnetic fields as compared to the experiment. This can be easily understood because spin decoherence microscopic mechanisms (apart from cotunnelling-like finite lifetimes) are not included in the calculations but are surely present in the experiments (see Ref.~\cite{Lan455} and supplementary section). To be more specific, finite-bias decoherence is also the cause of the experimental suppression of the $SU(4)_{B=0 T}$/$SU(2)_{B=10 T}$ peaks ratio observed in Fig. \ref{fg:H64C2KONDOCOND288mK0T10TA} (b). This discrepancy with theory, see Fig. \ref{fg:H64C2KONDOCOND288mK0T10TB} (a) and Fig. \ref{fg:H64C2KONDOCOND288mK0T10TB} (b) or Ref.~\cite{Shi195345}, has nothing to do with Kondo physics but with the fact that the experimental peak at $B$ = 10 T is less affected by decoherence processes. Overall, there is a very good agreement between the theoretical results and the experiments, as in both case the survival, even at very high magnetic fields, of a simple central ($V_b$ $\sim$ 0 mV) pure orbital Kondo related peak, is observed. This supports an interpretation of the experimental data in terms of an SU(4) to SU(2) crossover driven by magnetic field, where we tune the system between two distinct Kondo states. This behaviour is also in agreement with experiments in carbon nanotube quantum dots in the presence of a perpendicular magnetic field \cite{MakR241407}. Similarly to our case, the magnetic field in these experiments only couples to the spins and renders the orbital fluctuations unaffected. In conclusion, we have demonstrated a controlled $crossover$ between SU(4) and SU(2) Kondo states driven by magnetic field in a nanoscale Si transistor. The latter SU(2) Kondo effect originates {\em only} from quantum fluctuations in the orbital (valley) degree of freedom. We also perform theoretical calculations, including both spin and orbital Kondo physics as well as non-equilibrium effects, which confirm our experimental findings. Overall, the importance of the orbital degree of freedom in novel nano-scaled Si systems is confirmed, opening the way to the possible use of different symmetries that the donor orbitals can provide for innovative implementations of Si quantum-electronics \cite{Gre1057} such as for example, valley-based quantum computation \cite{Cul1}.

Acknowledgments: This work was supported by the EC FP7 FET-proactive NanoICT projects MOLOC (215750) and AFSiD (214989) and the Foundation for Fundamental Research on Matter (FOM). This research was also conducted within the ARC-CQC2T (project number CE110001027). R.A. and M.J.C. acknowledge funding from MICINN (Spain) through Grant No. FIS2009-08744. M.J.C. also acknowledges the Ram\'on y Cajal program, MICINN (Spain). The single dopant device was fabricated by N. Collaert and S. Biesemans at IMEC, Leuven.

\section{Supplementary Section: Magnetic Field Probing of an SU(4) Kondo Resonance in a Single Atom Transistor.}

\subsection{Magnetic Transport Data of the $D^{0}$ and $D^{-}$ peaks}

\begin{figure}
\begin{center}
\resizebox{8 cm}{!}{\includegraphics{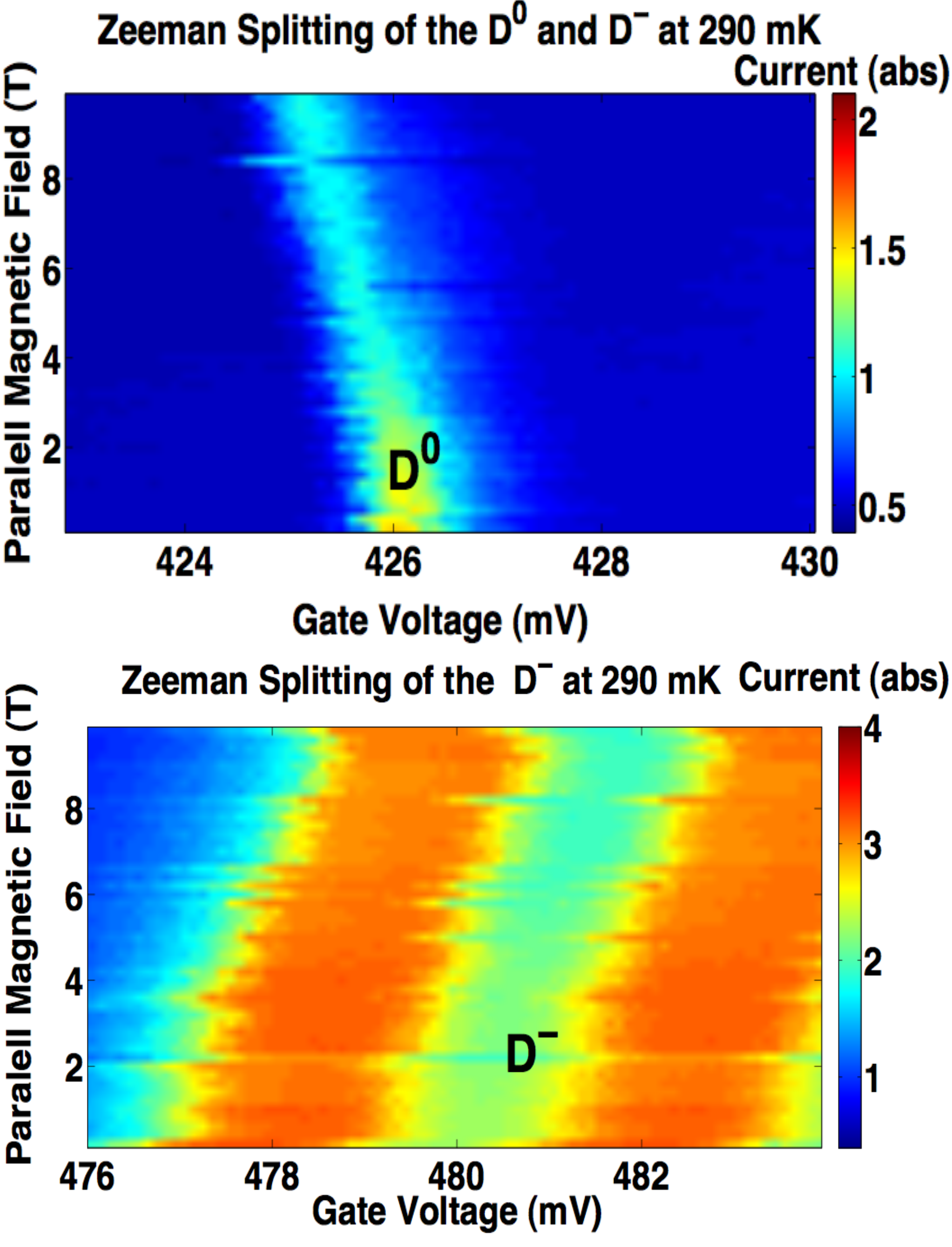}} 
\caption{Magnetic transport data of the $D^{0}$ and $D^{-}$ peaks for the device discussed in the main manuscript.}
\label{fg:KONDOTat450mVand10T}
\end{center}
\end{figure}

\begin{figure}
\begin{center}
\resizebox{7 cm}{!}{\includegraphics{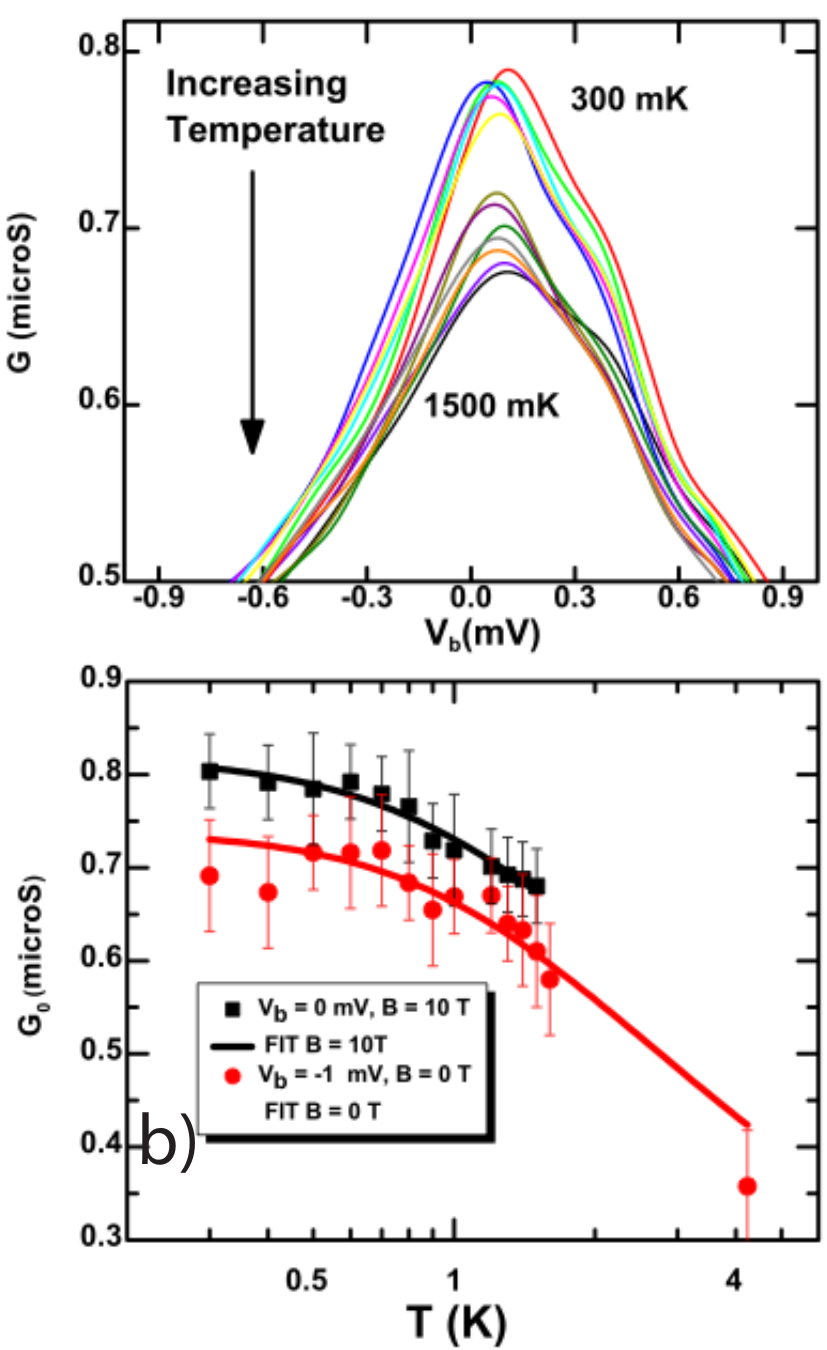}} 
\caption{a) Evolution with temperature of the central peak at $V_{g}\approx 450$ mV and for $B$ = 10 T between 0.3 K and 1.5 K. This situation is similar to the one theoretically described in Ref.~\cite{Tro125323}. b) Evolution in temperature of the peaks maximum at $B$ = 0 T (in red and with $T_{K}$ $\approx$ 8 K) and at $B$ = 10 T (in black and with $T_{K}$ $\approx$ 6 K). To obtain the $T_K$, we have fitted the data using the empirical law proposed by Goldhaber-Gordon \textit{et al} \cite{Gol5225}. Instrumental error bars compatibles with a 98 $\%$  confident band have been used.}
\label{fg:fig2supp}
\end{center}
\end{figure}

In Fig.~\ref{fg:KONDOTat450mVand10T} it is shown that, as expected for a single atom transistor \cite{Sel206805}, an odd/even effect in the magneto-transport spectroscopy data is observed. This consists of the Zeeman down shift  of the first electron resonance ($D^0$) and the Zeeman up shift of the second electron resonance ($D^-$) \cite{Sel206805}. The doubling of the $D^{-}$ peak is a feature sometimes observed in a single dopant system and can be associated with the presence of a nearby defect quickly switching its charge state (see Ref.~\cite{Lan455} and references therein). These results, together with the charging energy of 30-35 mV between the $D^0$ and the $D^-$ state in this sample (Fig.~\ref{fg:DH64ChipC2T300mKStabCONDBIS} (a) of the main text) and the identification of the excited states as in Ref.~\cite{Lan656}, confirm the single dopant nature of the Coulomb blockade transport observed in the device.

\subsection{Kondo Temperature for the Central Resonance at 0 T and at 10 T}

In Fig.~\ref{fg:fig2supp}, the evolution with temperature of the central resonance for $B$ = 0 T and $V_{g}$ = $450$ $mV$ (centred at $V_b$ = $0$ $mV$) and the evolution in temperature for both the $B$ = 0 T and $B$ = 10 T peaks are shown. A value of $T_{K}$ = 6 K is obtained for the $B$ = 10 T case, which is higher if compared to SU(2) spin Kondo temperatures observed for quantum dots (typically $\lesssim$ 1K) \cite{Lan455}. Furthermore, the point for $B$ = 0 T and $T$ = 4.2 K is considerably lower  if compared to the values in the 0.3 K-1.6 K range. The peak saturation observed for $\lesssim$ 1K in the $B$ = 0 T case, strongly supports the conclusion that, at $T$ $\sim$ 300 mK, our system is indeed in the zero temperature limit.

\begin{figure*} 
\includegraphics[width=0.45\textwidth,clip]{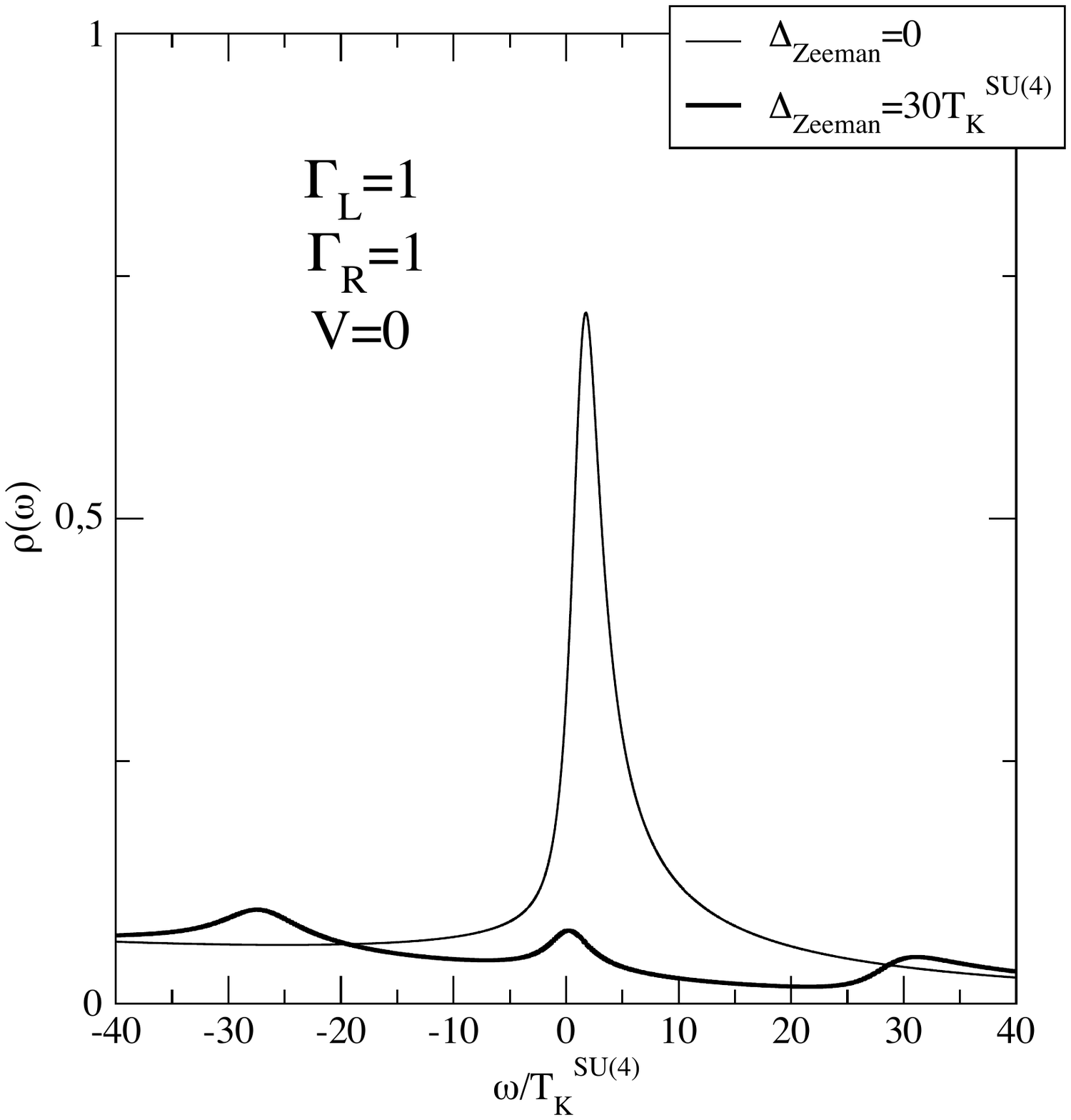}
\includegraphics[width=0.45\textwidth,clip]{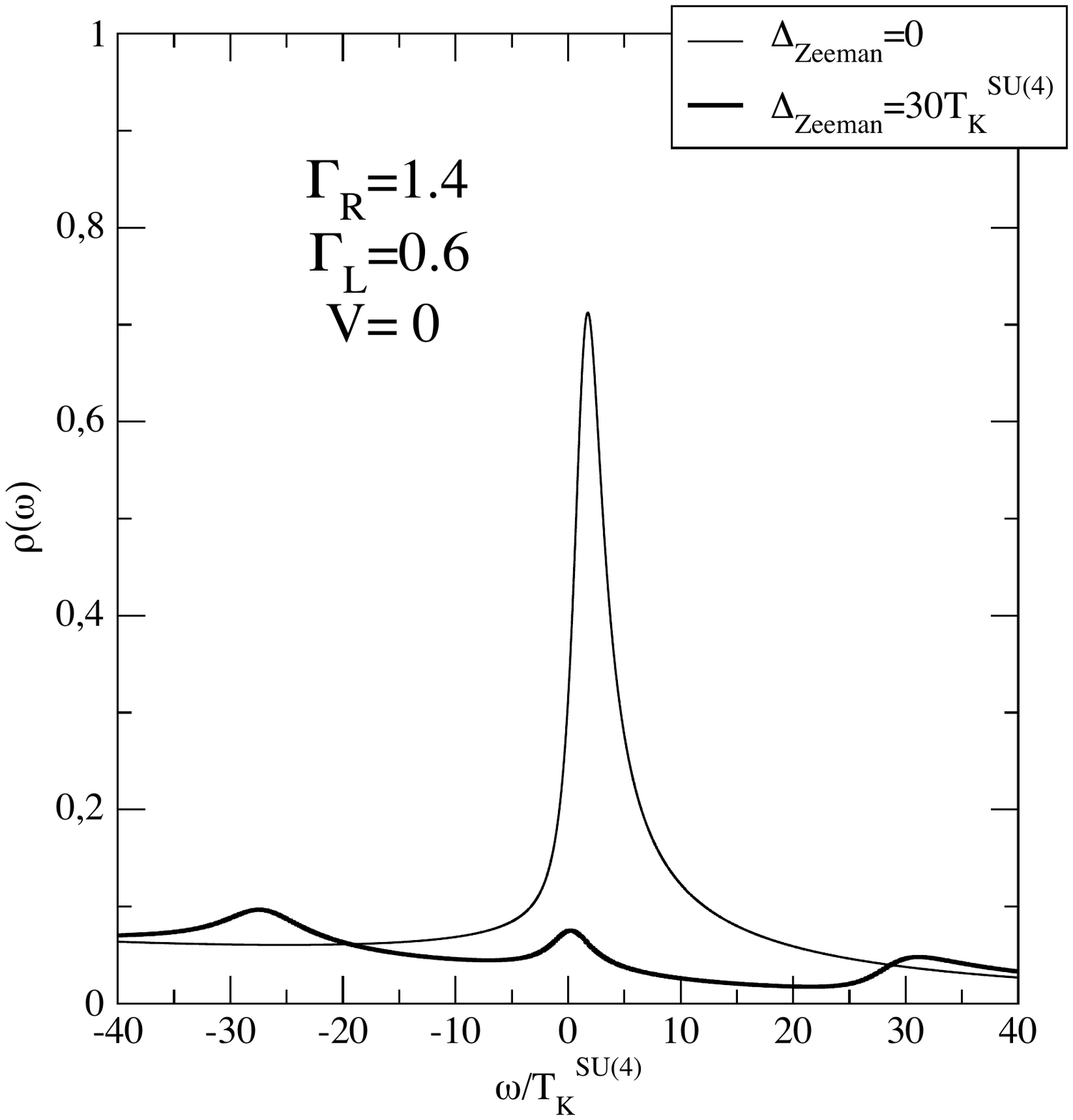}
\caption{Density of states for $V_{\rm bias}=0$ with symmetric (left) and asymmetric (right) couplings to the left $\Gamma_L$ and right $\Gamma_R$ reservoirs.
}
\label{fig:dosV0}
\end{figure*}

\subsection{Theory Model}

A single dopant in a FinFET can be described as a quantum dot (QD): The single electron state corresponds to the neutral donor $D^0$ while the two-electron state corresponds to a negatively charged donor $D^-$. Consider then a QD with two (nearly) degenerate localised orbitals (valleys) coupled to reservoirs. Hereafter we will denote this orbital quantum number by $\mu=1,2$. The dot is then described by the Hamiltonian 

\begin{eqnarray}
\label{su4vs2::eq:HD}
H_D = \sum_{\mu=1,2}\sum_{\sigma=\uparrow,\downarrow}
\varepsilon_{\mu\sigma} d_{\mu\sigma}^\dag d_{\mu\sigma}
+ \sum_{(\mu,\sigma)\neq(\mu',\sigma')} U_{\mu\mu'}n_{\mu\sigma}n_{\mu'\sigma'} \,,
\end{eqnarray}

where $\varepsilon_{\mu\sigma}$ is the single-particle energy level of the
localized state with orbital $\mu$ and spin $\sigma$, $d_{\mu\sigma}^\dag$
($d_{\mu\sigma}$) the fermion creation (annihilation) operator of the
state, \begin{math}
n_{\mu\sigma} = d_{\mu\sigma}^\dag d_{\mu\sigma}
\end{math}
the occupation, $U_{\mu\mu}$ ($\mu=1,2$) the intra-orbital Coulomb
interaction, and $U_{12}$ the inter-orbital Coulomb interaction.
The effect of the external magnetic field is to lift the spin degeneracy
of the single-particle energy levels.  The resulting splitting is denoted $\Delta_Z$ so that the
single-particle energy levels $\varepsilon_{\mu\sigma}$ have the form $\varepsilon_{\mu,\uparrow/\downarrow} = \varepsilon_0\pm(\Delta_Z/2)$.

On the other hand, as we argue in the main text, the valley (orbital) states are s-like \cite{Lan656} and are therefore not affected by the magnetic field.
The precise values of the Coulomb interactions $U_{\mu\mu'}$ depend on
the details of the system, but should be of the order of the
charging energy $E_{D^-}-2E_{D^0}\sim 36$ meV~\cite{U-PRB2010}.
We focus on the regime where the system
of the localized levels is occupied by a single electron
($\sum_{\mu\sigma}\langle{n_{\mu\sigma}}\rangle\approx 1$, quarter filling),
and the charging energy is much bigger than other energy scales {\bf (i.e. $U\rightarrow \infty$)}. In
this regime the Hamiltonian in Eq.~(\ref{su4vs2::eq:HD}) suffices
to describe all relevant physics.

Kondo physics arises as a result of the interplay between the strong
correlation in the dot and the coupling of the localized electrons with
the itinerant electrons in conduction bands. In our case, these are described as two leads ($\alpha=L$ and $R$) which are treated as non-interacting gases of
fermions

\begin{equation}
\label{su4vs2::eq:HC1}
H_\alpha =
\sum_k\sum_{\mu=1,2}\sum_\sigma\varepsilon_{\alpha k \mu}\,
a_{\alpha k\mu\sigma}^\dag a_{\alpha k\mu\sigma} \,.
\end{equation}

When the leads have the same symmetry as the dot, the orbital quantum number $\mu$ in the
leads is identical to the orbital quantum number in the dot and will be preserved over the tunnelling of electrons from the dot to the leads and vice versa. This is what we expect to have in our system with heavily doped Si as source and drain and a quantum dot that results from the hybridisation of the donor hydrogen-like state with the quantum well state formed in the Si channel~\cite{Lan656}. This  situation is described by the tunnelling Hamiltonian

\begin{figure*} 
\includegraphics[width=0.45\textwidth,clip]{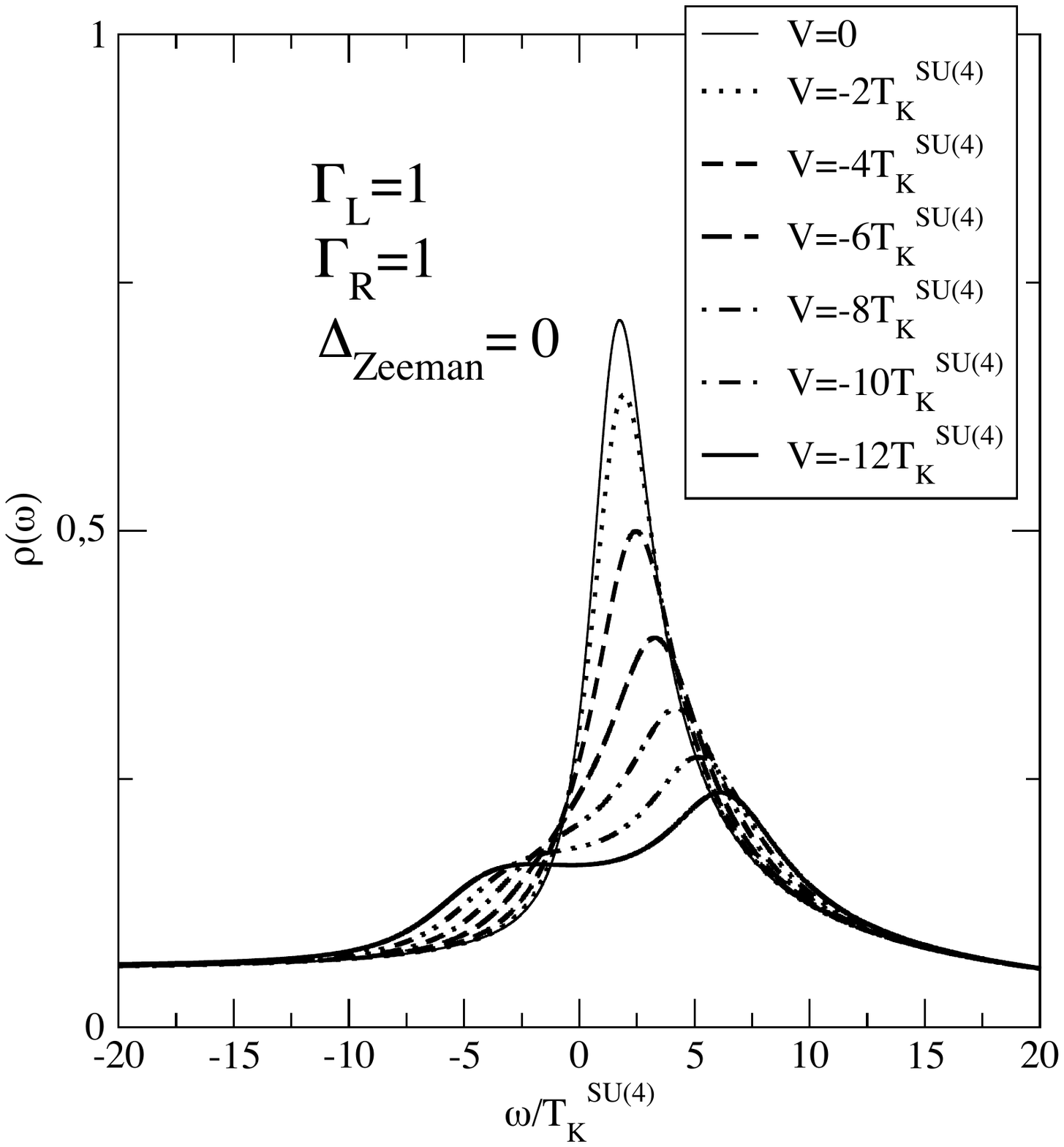}
\includegraphics[width=0.45\textwidth,clip]{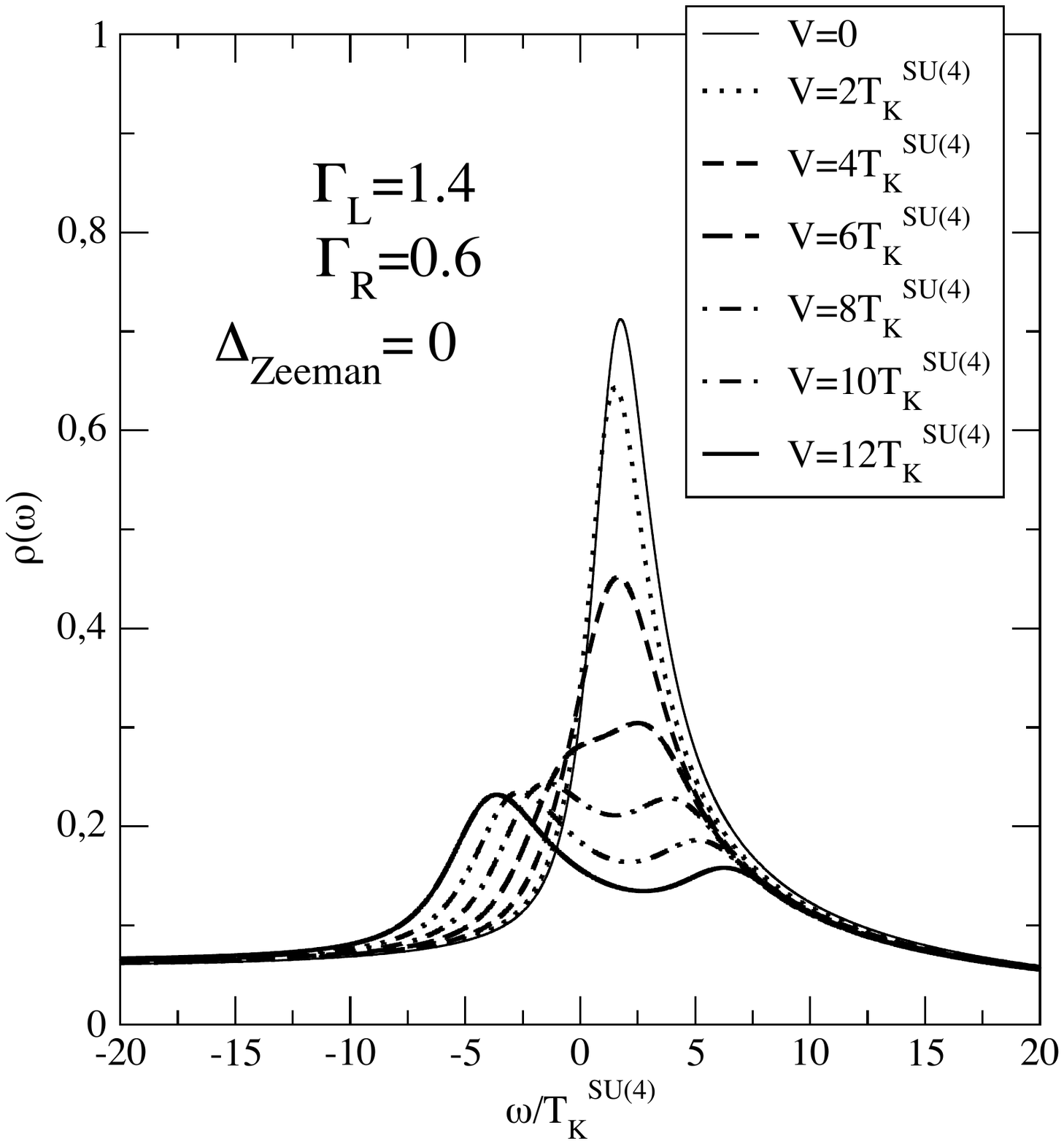}
\caption{Density of states for zero Zeeman spliting and different values of $V_{\rm bias}$ with symmetric (left) and asymmetric (right) couplings to the left $\Gamma_L$ and right $\Gamma_R$ reservoirs. A finite bias splits the Kondo peak in 2.
}
\label{fig:dosZeeman0}
\end{figure*}

\begin{figure*} 
\includegraphics[width=0.45\textwidth,clip]{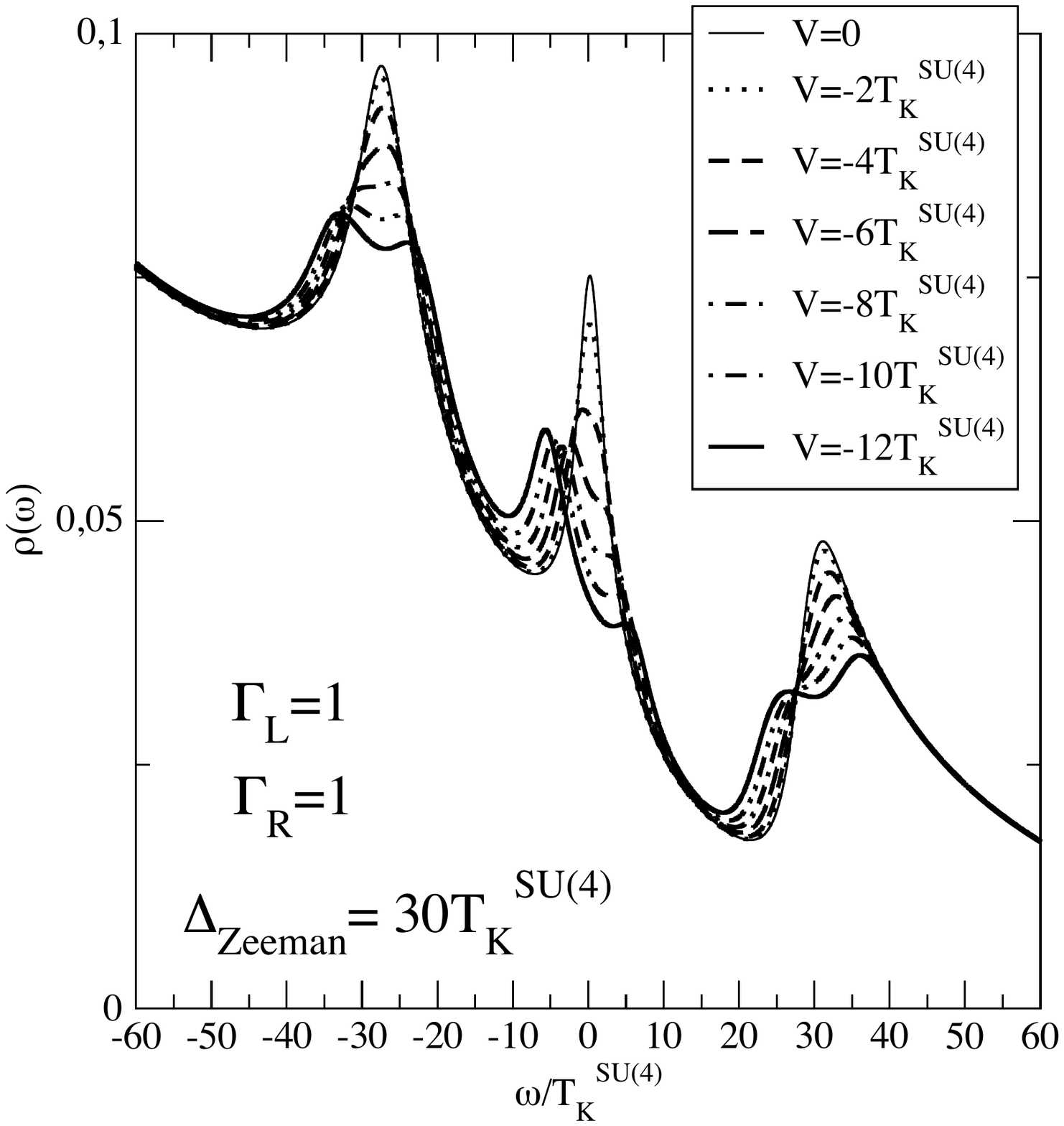}
\includegraphics[width=0.45\textwidth,clip]{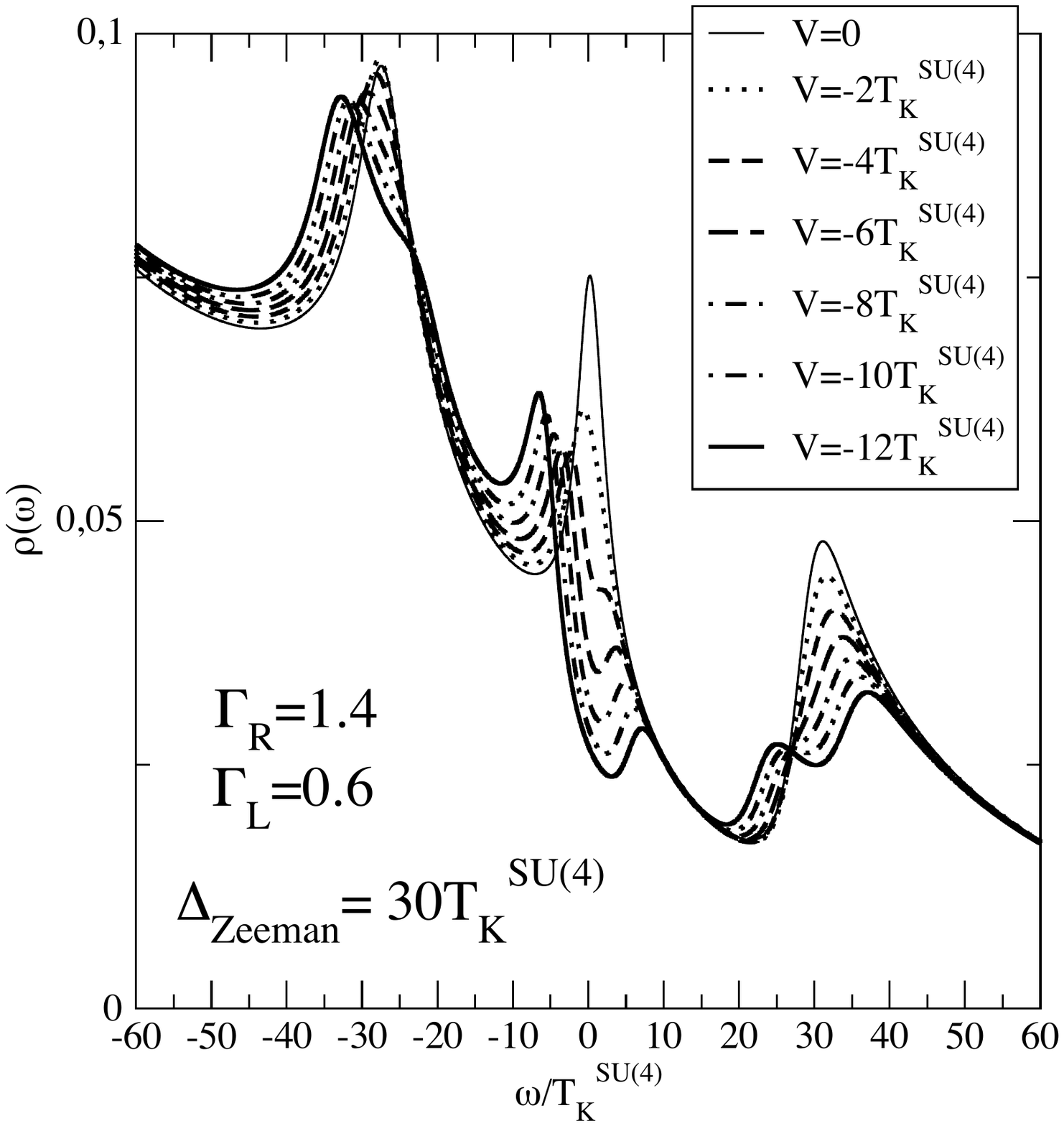}
\caption{Density of states for finite Zeeman splitting and different values of $V_{\rm bias}$ with symmetric (left) and asymmetric (right) couplings to the left $\Gamma_L$ and right $\Gamma_R$ reservoirs.
}
\label{fig:dosZeeman30}
\end{figure*}

\begin{equation}
\label{su4vs2::eq:HT1}
H_T =
\sum_{\alpha k\mu\sigma}
\left(V_{\alpha k\mu \sigma}
  a_{\alpha k\mu\sigma}^\dag d_{\mu\sigma}
  + h.c.\right) \,.
\end{equation}

The total Hamiltonian is then given by \begin{math} H = H_L + H_R + H_T + H_D
\end{math} \,. For simplicity, we ignore the $k$- and $\sigma$-dependence of the tunneling amplitudes. Therefore, we consider a simplified model with $V_{\alpha k\mu
\sigma}=V_{\alpha}/\sqrt{2}$ which defines the widths $ \Gamma_\alpha = \pi\rho_0 |V_\alpha|^2$,
where $\rho_0 $ is the density of states in the reservoirs. 

\subsection{Non-crossing approximation method} 
Now we write the physical fermionic operator as a combination of a
pseudo-fermion and a boson operator as follows: $d_{\mu\sigma}=b^\dagger
f_{\mu\sigma}$ where $f_{\mu,\sigma}$ is the pseudo-fermion which
annihilates one ''occupied state'' in the $\mu$th localised orbital with spin $\sigma$,
and $b^\dagger$ is a boson operator which creates an ''empty
state''. Quite generally the intra(inter) Coulomb interaction is very
large and we can safely take the limit of $U\rightarrow \infty$.  This
fact enforces the constraint

\begin{math}
\sum_{\mu\sigma}f_{\mu\sigma}^{\dag}f_{\mu\sigma} + b^\dag b=1
\end{math},
that prevents the accommodation of two
electrons at the same time in either the same orbital or different
orbitals.  This constraint is treated with a Lagrange multiplier.

\begin{equation}
\label{hamiltonian1}
H_\mathrm{SB}
= \sum_{k,\sigma,\mu}\varepsilon_{k_{\mu}}c_{k_\mu,\sigma}^\dag
c_{k_\mu,\sigma}+ \sum_{\sigma,\mu}\varepsilon_{0,\sigma}
f_{\mu,\sigma}^\dag f_{\mu,\sigma}
+ \frac{\overline{V}_\mu}{\sqrt{N}}\sum_{k,\sigma, \mu}
\, \left(c_{k_{\mu},\sigma}^\dag b^\dag f_{\mu,\sigma} + h.c.\right)
+  \lambda \left(\sum_{\mu,\sigma} f_{\mu,\sigma}^\dag f_{\mu,\sigma}
  + b^\dag b - 1\right) \,.
 \end{equation}

Notice that we have rescaled the tunneling amplitudes $
V_{\mu} \to \overline{V}_{\mu}\sqrt{N}$ according to the spirit of a $1/N$-expansion ($N$ is the total
degeneracy of the localized orbital).

Our next task is to solve this Hamiltonian, which is rather
complicated due to the presence of the three operators in the
tunnelling part and the constrain. In order to do this we employ
the so-called Non-Crossing approximation (NCA) \cite{NCAneq1,NCAneq2,NCAneq3}. Without entering into much detail of the theory, we just mention that the boson fields in Eq.~(\ref{hamiltonian1}) are treated as fluctuating operators such that both thermal and charge fluctuations are included in a self-consistent manner. In particular, one has to derive self-consistent equations-of-motion for the time-ordered double-time Green's function (sub-indexes are omitted here):

\begin{eqnarray}
iG(t,t')&\equiv&\langle T_c f(t)f^\dagger(t')\rangle\nonumber\,,\\
iB(t,t')&\equiv&\langle T_c b(t)b^\dagger(t')\rangle,
\end{eqnarray}
or in terms of their analytic pieces:
\begin{eqnarray}
iG(t,t')&=&G^{>}(t,t')\theta(t-t')-G^{<}(t,t')\theta(t'-t)\,,\nonumber\\
iB(t,t')&=&B^{>}(t,t')\theta(t-t')+B^{<}(t,t')\theta(t'-t);
\end{eqnarray}

A rigorous and well established way to derive these
equations-of-motion was first introduced by Kadanoff and

\begin{figure*} 
\includegraphics[width=0.45\textwidth,clip]{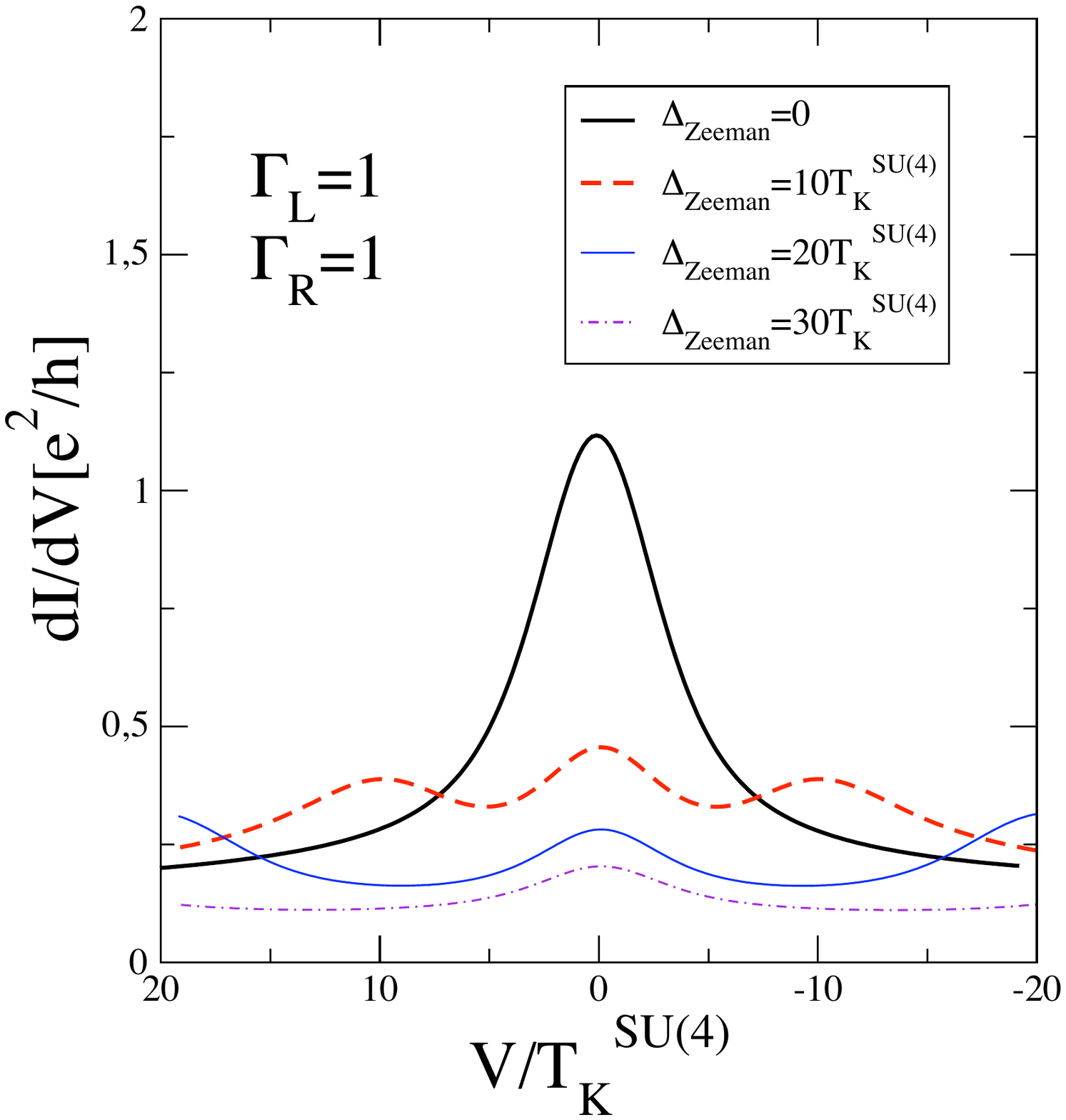}
\includegraphics[width=0.45\textwidth,clip]{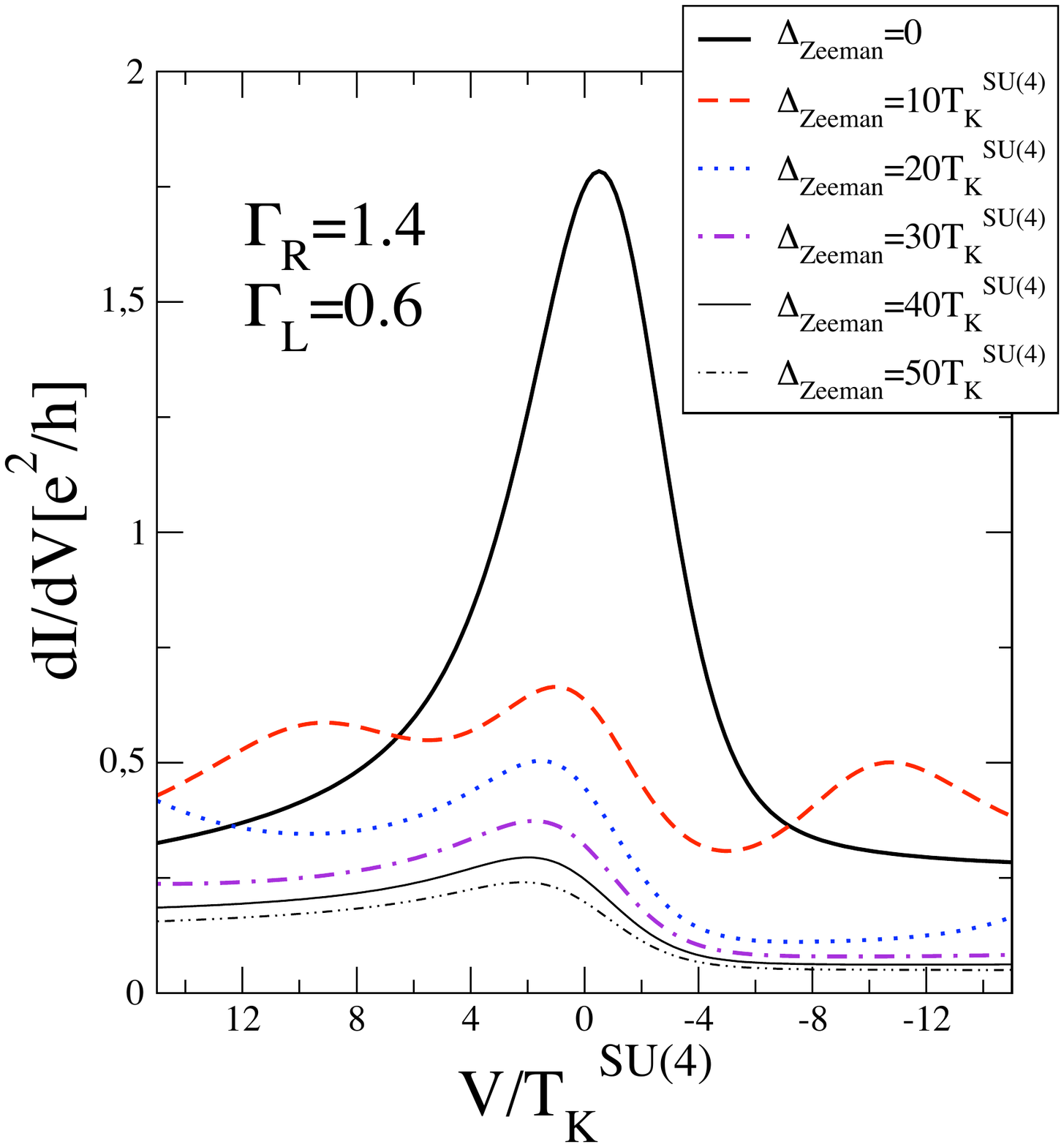}
\caption{Differential conductance for symmetric (left) and asymmetric (right) contacts for different magnetic fields. The conductance is given in units of the quantum of conductance $e^2/h$ (which corresponds to 38.74$\mu S$, as presented in the main text).
}
\label{fig:didv}
\end{figure*}

\noindent Baym \cite{Kadan}, and has been related to other non-equilibrium
methods (like the Keldysh method) by Langreth, see
Ref.~\cite{lan76} for a review. Here, we just show numerical
results of the NCA equations for our problem and refer the
interested reader to Refs.~\cite{NCAneq1,NCAneq2,NCAneq3,ram03} for details. In particular, the density of states is given by

\begin{equation}
\rho(\omega)=-\frac{1}{\pi}\sum_{\mu,\sigma}
\mathrm{Im}[A^{r}_{\mu\sigma}(\omega)],
\end{equation}

where $A^{r}_{\mu\sigma}(\varepsilon)$ is the Fourier transform of the retarded Green's function
$A^{r}_{\mu\sigma}(t) = G_{\mu,\sigma}^{r}(t)B^{<}(-t)-G^{<}_{\mu\sigma}(t) B^{a}(-t)$. Following Meir and Wingreen in Ref. \cite{Meir92a}, the current is given by:

\begin{equation}
I_{\alpha\in \{ L,R \}}=-\frac{2e}{h}\sum_{\mu,\sigma}\int d\epsilon
\Gamma_\alpha(\epsilon)[2Im A^r_{\mu\sigma}(\epsilon)f_\alpha(\epsilon)
+A^<_{\mu\sigma}(\epsilon)].\nonumber\\
\end{equation}

with $A^<_{\mu\sigma}(\epsilon)$ the Fourier transform of $A^{<}_{\mu\sigma}(t) = iG_{\mu,\sigma}^{<}(t)[B^{r}(-t)-B^{a}(-t)]$ and $f_\alpha(\epsilon)=\frac{1}{1+e^{\frac{(\epsilon-\mu_\alpha)}{kT}}}$ the Fermi-Dirac function at each reservoir held at a chemical potential $\mu_\alpha$ such that the applied bias voltage is defined as $eV=\mu_R-\mu_L$. 

\subsection{Results}
In the following, we present results for the density of states and the differential conductance $\frac{dI}{dV}=\frac{dI_L}{dV}=-\frac{dI_R}{dV}$. Fig.~\ref{fig:dosV0} shows the density of states (DOS) for {\bf zero bias} for symmetric and asymmetric couplings to the reservoirs. For $B=0$ (zero Zeeman splitting), the peak on the DOS is away from zero frequency while for a finite Zeeman splitting $B \neq 0$ the central resonance [due to the orbital degree of freedom] shifts to $\omega=0$ (shown in the Figure for $\Delta_{\rm Zeeman}=30T_K^{SU(4)}$). The results for symmetric and asymmetric couplings are exactly the same because $\Gamma_L+\Gamma_R$ is the same in both cases.

The behaviour is very different for finite bias. Fig.~\ref{fig:dosZeeman0} shows the DOS for $B=0$ and different values of the bias voltage. First, note that eventually, for a certain value of $V$, the DOS peak splits in two. $V$ is applied such that the left reservoir has $\mu_L=V/2$ and the right one $\mu_R=-V/2$ with respect to the Fermi level of the dot. The splitting in the DOS is due to the coupling to the two reservoirs displaced with respect to the Fermi energy. The splitting does not occur symmetrically because the dot {\em sees} the two reservoirs differently, even in the case of symmetric couplings $\Gamma_R=\Gamma_L$. If a magnetic field is also applied, a Zeeman splitting occurs (see Fig.~\ref{fig:dosZeeman30}) and, in turn, each new peak splits due to the finite bias.

Let us turn now to the $dI/dV$ curves. For symmetric contacts ($\Gamma_L=\Gamma_R=1$) the differential conductance peak is always centred at zero bias (see Fig.~\ref{fig:didv}(left)). A magnetic field produces a Zeeman splitting and the peak that remains at $V=0$ is due to the orbital degree of freedom. Remember that the peak of the DOS for the symmetric case and $B=0$ is away from $\omega=0$ (Fig.~\ref{fig:dosV0} (left)), however, the $dI/dV$ is peaked at zero bias. On the other hand, for the asymmetric case (shown in Fig.~\ref{fig:didv} (right)) the differential conductance peak occurs at finite positive bias for $B=0$. Applying a magnetic field produces a Zeeman splitting and a shift towards negative values of the bias of the central SU(2) peak. The position of this central peak seems to saturate for a sufficiently large value of the magnetic field. This can be understood from the density of states: as the magnetic field splits the Kondo resonance, owing to the Zeeman splitting of the spin sector, more spectral weight is displaced towards lower energies (close to the Fermi energy), see Fig.~\ref{fig:dosZeeman30}. When the coupling to the leads is asymmetric, this is reflected as a Kondo peak in the differential conductance that moves towards lower bias voltages (near zero bias). The peak saturates when spin fluctuations are frozen by the magnetic field. This asymmetric coupling situation is the one discussed in the main text to show agreement with the experimental data.

\end{document}